\documentclass[manuscript]{acmart}

\usepackage{gensymb}
\usepackage{xcolor}
\usepackage{float}
\floatstyle{plaintop}
\restylefloat{table}
 \usepackage{amsmath}
 \usepackage{xr}
 \usepackage{textcomp}
 \usepackage{graphicx}
 \usepackage{tabularx}
 \usepackage{colortbl}
 \usepackage{comment}
 \usepackage{multirow}
 \usepackage{booktabs}
\usepackage{hyperref}
\usepackage{hyperxmp}
\DeclareUnicodeCharacter{202F}{\,}

\AtBeginDocument{%
  \providecommand\BibTeX{{%
    \normalfont B\kern-0.5em{\scshape i\kern-0.25em b}\kern-0.8em\TeX}}}

\begin{document}

\title{Free-Will vs Free-Wheel: Understanding Community Accessibility Requirements of Wheelchair Users through Interviews, Participatory Action, and Modeling}

\author{Hanna Noyce}
\affiliation{%
  \institution{University of Wisconsin Milwaukee}
  \city{Milwaukee}
  \state{Wisconsin}
  \country{USA}}

\author{Emily Olejniczak}
\affiliation{%
  \institution{University of Wisconsin Milwaukee}
  \city{Milwaukee}
  \country{USA}
}

\author{Vaskar Raychoudhury}
\affiliation{%
 \institution{Miami University}
 \streetaddress{510 E. High Street}
 \city{Oxford}
 \state{Ohio}
 \country{USA}}

\author{Roger O. Smith}
\affiliation{%
  \institution{University of Wisconsin Milwaukee}
  \city{Milwaukee}
  \state{Wisconsin}
  \country{USA}}



\author{Md Osman Gani}
\affiliation{%
  \institution{University of Maryland Baltimore County}
  \city{Baltimore}
  \country{USA}}


\begin{abstract}
Community participation is an important aspect of an individual's physical and mental well-being. 
This participation is often limited for persons with disabilities, especially those with ambulatory impairments due to the inability to optimally navigate the community. Accessibility is a multi-faceted problem and varies from person to person. Moreover, it depends on various personal and environmental factors. Despite significant research conducted to understand challenges faced by wheelchair users, developing an accessibility model for wheelchair users by identifying various characteristic features has not been thoroughly studied. 
In this research, we propose a three-dimensional model of accessibility and validate it through in-depth qualitative analysis involving semi-structured interviews and participatory action research. The outcomes of our studies validated many of our hypotheses about community access for wheelchair users and identified a need for more accessible path planning tools and resources. Overall, this research strengthened our three-dimensional User-Wheelchair-Environment model of accessibility. 

\end{abstract}





\maketitle

\section{Introduction}


Individuals with mobility impairments often face immense difficulties when trying to access the community. Specifically, users of wheeled mobility-aids such as  wheelchairs or mobility scooters are often hindered by innumerable barriers in the built and natural environments which may seem insignificant to able-bodied individuals. A single step or a short portion of a path with a steep slope can thwart wheelchair users and even leave them stranded. There are also various temporary barriers caused by constructions, fallen trees, or discarded furniture which present themselves unexpectedly and leave regularly used accessible paths non-navigable. Problems are far more pronounced in an unfamiliar environment such as a university campus or a tourist destination that attracts numerous visitors on a regular basis. Wheelchair users need to spend considerable time researching the accessibility of the indoor or outdoor spot they plan to visit through various data sources such as maps, crowd-sourced information, and reviews. Existing mapping systems fail to provide crucial accessibility information to the wheeled mobility-aid users which significantly limits their community participation. 

\begin{figure*}
  \centering
  \includegraphics[width=0.99\linewidth]{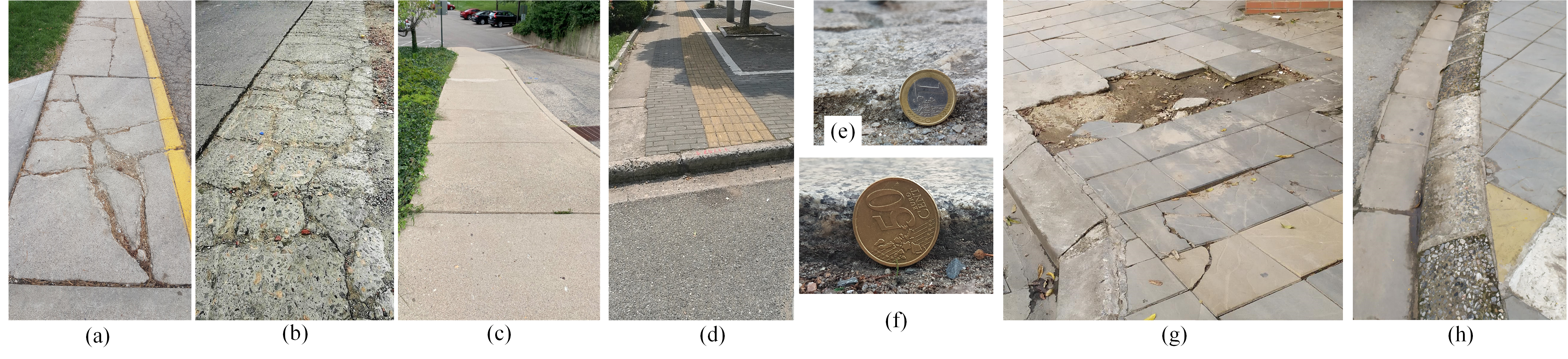}
  \caption{(a) Broken Sidewalk, USA, (b) Narrow \& Broken Side-road, China, (c) Steep Slope (9-12$^{\circ}$), USA, (d) Curb with no Ramp, China, (e) Cobblestone Surface, Germany, (f) Uneven Sidewalk, Germany, (g) Broken Sidewalk, Vietnam, (h) Steep Curb Ramp, Vietnam}
  \label{fig:barriers}
  \Description{The eight subfigures of this Figure shows inaccessible road features from different parts of the world, which are insurmountable by wheelchair users. The first figure shows a broken sidewalk in Oxford, OH, USA with deep lateral and longitudinal grooves for a continuous 8 meters of the sidewalk, which is not possible to pass by all capability levels of wheelchair users. The second figure shows similar sidewalk features in Zhejiang, China. The third figure captures a steep sidewalk with a 9 to 12-degree slope in Oxford, OH, USA. The fourth figure has a curb without any access ramp or curb cut in Zhejiang, China. The fifth figure shows cobblestone surfaces in Mannheim, Germany with deep grooves of 2 cm or more in between. In fact, this is considered to be the most difficult and painful surface to navigate for any wheelchair user. The sixth figure shows sudden surface changes in sidewalks in Dresden, Germany with sudden deep and uneven gaps of 1.3 to 2 cms in between. The seventh figure shows a sidewalk with broken square tiles for a large part of the surface found in Ha Noi, Vietnam. The eighth and final figure is also from Ha Noi, Vietnam, and has a steep 60-degree curb ramp which is treacherous.}
 \end{figure*}

Matthews, et al. \cite{matthews2003modelling} conducted a survey of 27 wheelchair users in 2003 to understand their navigational experiences and the outcome identified several barriers including the absence of curb ramps, poorly laid surfaces, weather (rain and snow), etc. To facilitate the accessibility of public spaces for all users different regulations have been introduced in different countries across the globe \cite{ADA, DIN_EN_18024}. Americans with Disabilities Act \cite{ADA} was introduced in 1990 to ensure equality of access to public areas. However, even after 30 years of introducing ADA regulations, accessibility needs are still not fully met which is evident from a more recent (2020) survey on wayfinding \cite{gupta2020towards} where wheelchair users are still concerned about ``ADA-noncompliant ramps that are too steep or texturally uneven". Fig.~\ref{fig:barriers} portrays examples of various barriers commonly found throughout the built environment. Our research team has identified those barriers from various places across  the globe. 

Unnavigable physical barriers deter independent excursions and future community participation of wheelchair users and create psychological barriers caused by trip delay, cancellation, and exhaustion. These negative experiences may result in withdrawal from community participation and outings, especially for new wheelchair users and older adults. To address this issue, we plan to develop a path planning tool based on a crowd-sourced accessibility map with marked barriers and facilities for accessible routing and navigation. Using this map, the path planning tool will generate personalized routes customized to the need of each wheelchair user, their mobility aid, and various environmental factors.

With around 4 million wheelchair users in the USA \cite{steinmetz2006americans}, wheelchairs are the most commonly used mobility aid for individuals with ambulatory disabilities. Based on an extensive literature survey combined with our everyday experiences, we figured out that accessibility changes as a complex function of the users (U), their wheelchairs (W), and the environment (E). In this research, we plan to understand the intricate relations between U, W, and E and the various parameters associated with them and develop an \textit{accessibility model}. We have proposed a set of hypotheses relating U, W, and E and went on to conduct extensive qualitative research to accept or reject those hypotheses. Our qualitative data collection is based on a two-pronged approach involving \textit{user interviews} and \textit{action research}. 
We interviewed 51 stakeholders who are either wheelchair users or individuals who have a role in personally or professionally supporting wheelchair users. An in-depth thematic analysis of the interview outcomes points to some distinct findings which are mostly in congruence with our initial hypotheses. Identified themes that were not hypothesized helped us to improve the proposed accessibility model. Further, we conducted Participatory Action Research (PAR) to incorporate feedback from experts regarding our proposed hypotheses and the accessibility model. PAR group members were domain experts on wheelchair use and accessibility and helped us to improve our initial hypotheses with critical analysis and meaningful insights. Combining the interview and PAR outcomes, we finalized a three-dimensional model of accessibility (3D-UWM) considering various factors that might affect accessible navigation. Compared with some well-known previous models, namely Nagi Model \cite{nagi1976epidemiology}, IOM model \cite{pope1997enabling}, and PEO model \cite{law1996person}, our proposed 3D-UWM model has been empirically evaluated using data collected through qualitative research. Some quantitative data collected using inertial sensing modules \cite{gani-accepted-a} show that our proposed model is also practically feasible. In summary, we made the following four contributions through this paper:

\begin{itemize}
    \item Identified three unique aspects of accessibility - Users (U), Wheelchairs (W), and the Environment (E) and listed various parameters associated with them. Also, identified a set of hypotheses involving U, W, and E and their interactions to build a full-scale model of accessibility.
    \item Conducted interviews of 51 stakeholders to qualitatively evaluate the proposed hypotheses. A thematic analysis bolsters our accessibility model with new findings as well as by confirming our hypotheses.
    \item Conducted Participatory Action Research (PAR) involving 12 stakeholders and collected useful insights into the various issues associated with the proposed hypotheses. 
    \item Finally, the outcome of the qualitative analyses of the interviews and the action research helped us to consolidate our accessibility model into a three-dimensional User-Wheelchair-Environment (3D-UWE) model which we will verify through further quantitative analysis.
\end{itemize}


\section{Related works}

Global improvement in healthcare and well-being combined with the reduced birth rate in many countries has resulted in significant growth in the elderly population \cite{united2017department} worldwide. Moreover, there are more than 61 million adults in the USA with some form of disability among which 13.7\% suffers from ambulatory disabilities \cite{CDCDisability}. Community participation is non-trivial for persons with disabilities (PwDs) because of various barriers to accessibility. This adversely affects the social life of the PwDs and leads to mental dissatisfaction and gradual withdrawal from the community. 

To address this issue, researchers have studied the problem of accessibility for people with different disabilities. Various tools and technologies have been proposed apart from the qualitative evaluation of the problems. Accessibility for the visually impaired population has been thoroughly investigated in literature \cite{campbell2014s, hara2015improving, guerreiro2020virtual}. Also, researchers have conducted qualitative as well as quantitative studies of accessibility for persons with ambulatory disabilities \cite{hara2016design, karimi2013personalized, kasemsuppakorn2015understanding, menkens2011easywheel}. Various tools and techniques have been proposed to find routes through the built environment for independent navigation. Research in this domain can be grossly divided into \textit{user surveys} \cite{meyers2002barriers} of mobility-impaired individuals and \textit{spatial analysis} \cite{matthews2003modelling, beale2006mapping, kurihara2004use} of public places to identify the mobility aspects of users and accessibility barriers present in the built environment \cite{holone2007users, neis2008openrouteservice, neis2012towards, neis2011street, neis2014recent, volkel2008routecheckr}. Various data collection techniques (manual and automated \cite{bujari2012movement}) have been developed for collecting (single-user and crowd-sourced \cite{traunmueller2013introducing, holone2007users, zambonelli2011pervasive, bicocchi2013collective}) information on accessibility barriers and assigning accessibility scores to different features. Web-based and smartphone-based mobility assistants have been developed for routing and navigation of mobility-impaired persons \cite{kawamura2008mobile, umezu2013context, quercia2014shortest}.




Although there is ongoing research on community accessibility to maintain an active lifestyle for older adults \cite{colby2015projections} and for PwDs, overall there is a significant research gap in addressing issues associated with accessibility and activity engagement in older adult populations \cite{cunningham2004concepts} and of wheelchair users. Moreover, stakeholder perspectives regarding individuals with disabilities and older adults' community accessibility route needs are virtually non-existent. 

In this research, we focus on community accessibility for persons with ambulatory disabilities and older adults who require wheelchairs for their daily mobility needs. However, accessibility is a complex phenomenon involving interaction between users, their mobility aids, and the surrounding environment. E.g., a path segment with $6^{\circ}$ incline might be inaccessible to an 80-year-old female with spinal cord injury while in a manual wheelchair. But, the same path segment can be accessed using a power wheelchair by the same user. 

Research in rehabilitation science and engineering has aimed to unearth the complex interaction between user and environment for decades. In 1976, Nagi \cite{nagi1976epidemiology} presented the first disability model in which some kind of pathology triggers an impairment, which may result in a limitation in body function and thereby cause a disability. Nagi's model considered the effect of the environment in terms of family, community, and society and how they affect the disability of a person. The Institute of Medicine (IOM) model \cite{pope1991disability}, proposed in 1991, was derived directly from Nagi, defining disability as "a function of the interaction of the person with the environment". A new three-dimensional IOM model (Enablement-disablement Model) was introduced in 1997 \cite{pope1997enabling}, in which, a deeper interaction of the person and his/her environment is manifested. The environment factor includes the physical, social, and psychological components of the environment. Around the same time, the famous Person-Environment-Occupation (P-E-O) Model \cite{law1996person} was proposed to capture the dynamic relationship between people or users, their roles or occupations, and their current surroundings or environment, in which they live and work. The interaction of the three elements results in the occupational performance of users. During the lifetime of an individual or community, the overlaps of the three components differ in size, due to different factors and their mutual interactions and produce different occupational performances. 

The modified IMO and the P-E-O models were transformative in the sense that, the earlier models were more people-centric with less attention to their interactions with the community and the environment. However, empirical analysis and stakeholder perspectives that establish the relationships between persons with disabilities (PwDs) and their surrounding environment, using actual data collected from the environment, are nearly absent. To explore this issue, our research group engaged in the National Science Foundation’s (NSF) Innovative Corps (I-CORPs) program \cite{icorps}, with interviews identifying stakeholder perspectives in the customer discovery process. The study interviews assessed the major barriers to accessibility for wheelchair users and if there exists a need for an accessible routing information application among wheelchair users, which included a significant number of older adult stakeholders.

We also involved various experts in a participatory action research (PAR) to understand and to verify the issues and needs raised by the interviewees. Overall, this is a unique qualitative approach to building a path accessibility model for wheelchair users involving various personal and environmental factors.

\section{Accessibility Model and Model Evaluation}


Based on the literature and our prior work, we hypothesize that: accessibility for wheelchair users depends on three unique aspects involving (1) the user, (2) the type of their wheelchair, and (3) the types of the environment. Each of these aspects has various associated parameters that interact and contribute to the overall fluency of the community outing experienced by a wheelchair user. We have analyzed and identified multiple such parameters and presented them in Table~\ref{tab:Characteristics}. However, the list is not comprehensive. In order to identify more parameters and to establish the complex relationship between the U, W, and E factors leading to a comprehensive accessibility model we plan to conduct a thorough qualitative evaluation.

\begin{table}[htbp]
\caption{Various Parameters Pertaining to Users, their Wheelchairs, and the Physical Environment.}
\label{tab:Characteristics}
\centering
\scalebox{0.8}{
\begin{tabular}{lll}
\toprule
\textbf{User Characteristics} &
\textbf{Wheelchair Characteristics} &
\textbf{Environment Characteristics} \\
\midrule
1. Type of Disability &
1. Type (manual, power, power-assist) &
1. Time \\

\begin{tabular}[c]{@{}l@{}}
2. Degree of disability (using standard \\
scales, like ASIA scale for SCI)
\end{tabular} &
2. Material (Steel, Titanium, …) &
2. Weather \\

3. Age &
3. Frame Type (rigid, flexible) &
3. Obstacles \\

4. Weight &
\begin{tabular}[c]{@{}l@{}}
4. Number of wheel pairs \\
(including main and casters)
\end{tabular} &
4. Width of surface \\

5. Height &
\begin{tabular}[c]{@{}l@{}}
5. Drive Type (rear-wheel, \\
front-wheel, mid-wheel)
\end{tabular} &
5. Type of surface \\

6. Gender &
6. Seat to floor height &
6. Presence/absence of curb ramp \\

7. Purpose of travel (urgent, non-urgent) &
7. Weight capacity &
\begin{tabular}[c]{@{}l@{}}
7. Presence/absence of pedestrian \\
crosswalk with or without signal
\end{tabular} \\

8. ……… &
8. Seat width &
8. ……… \\

&
9. Chair width &
\\

&
10. ……… &
\\
\bottomrule
\end{tabular}
}
\end{table}

\begin{figure}
\begin{center}
  \includegraphics[width=0.85\textwidth]{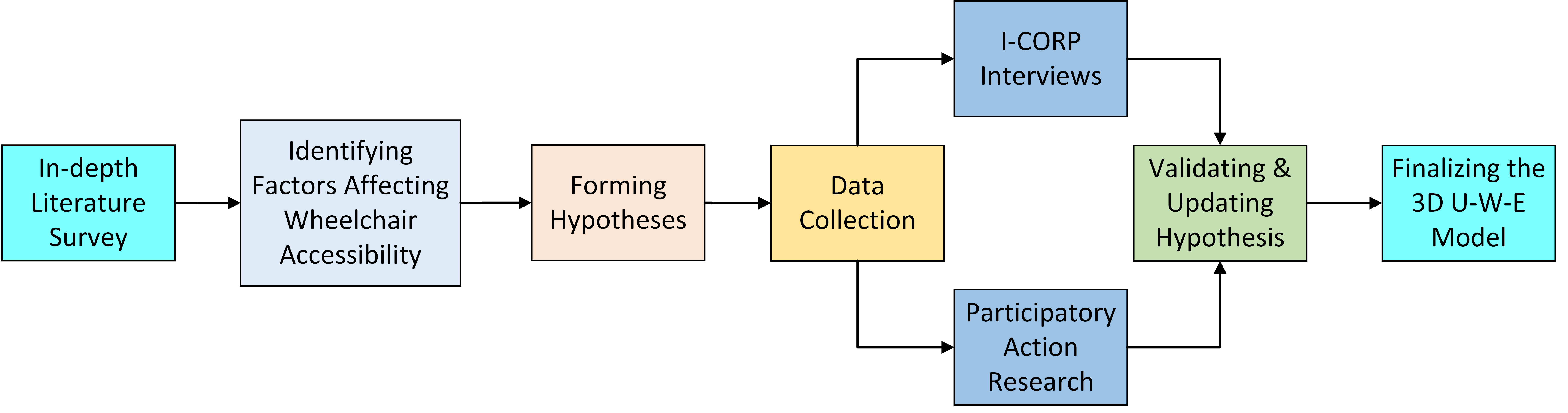}
    \end{center}
    \caption{Flowchart of Developing a Three Dimensional Accessibility Model}
    \label{fig:flow}
    \Description{The graphic depicted in this figure is a flowchart describing the development of a three-dimensional accessibility model through this research. The flowchart is a series of 8 boxes beginning with “in-depth literature survey”, followed by “identifying factors affecting wheelchair accessibility” “forming hypothesis”, and “data collection”. The chart then splits into a top box of “I-Corps interviews” and a bottom of “participatory action research”, both of which lead back to the box “validating & updating hypothesis. The chart ends with the final goal of the project which is “finalizing the 3D U-W-E Model”.}
\end{figure}

\subsection{Model Evaluation}

To develop and validate the accessibility model we followed an approach as described in the flowchart in Fig.~\ref{fig:flow}. We initially form some hypotheses based on the known interaction patterns of the U, W, and E factors. We describe our hypotheses in the following subsection. Once our hypotheses are ready, we started collecting data to evaluate those. Two independent approaches were conducted for the evaluation. The first method involved user interviews using a semi-structured approach while the second method used participatory action research to elicit responses from the stakeholders who know and can critically analyze the problem of accessible wheelchair navigation.

\begin{figure}
\begin{center}
\includegraphics[width=0.8\textwidth]{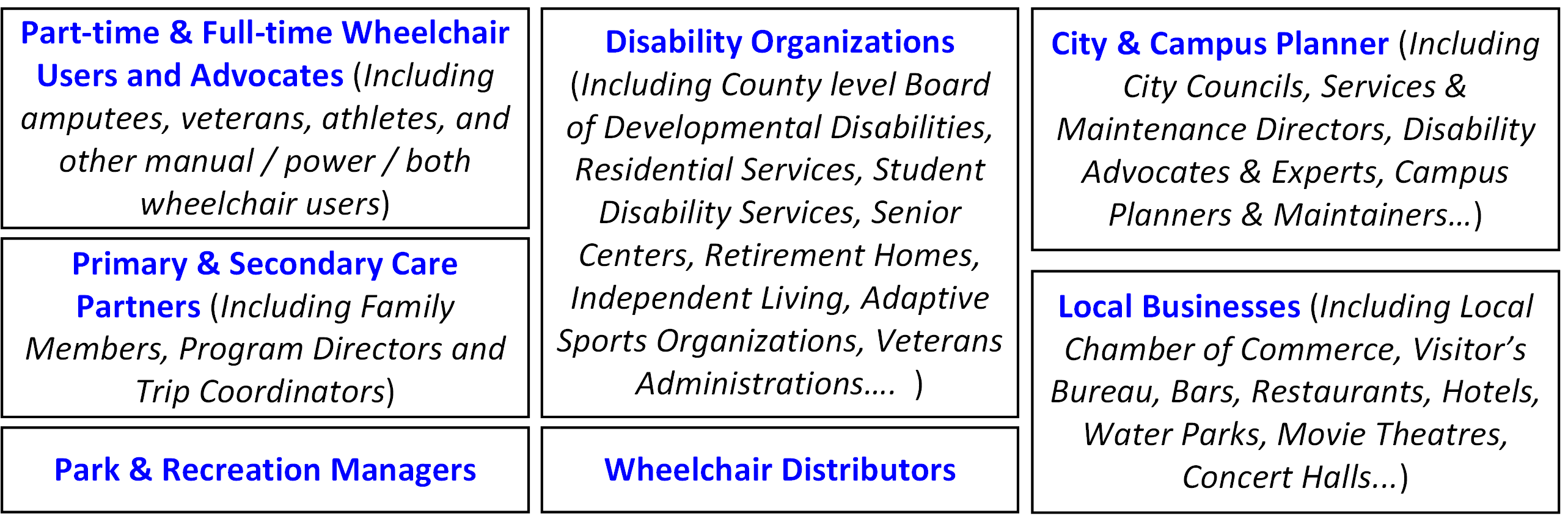}
    \end{center}
    \caption{Our Interview Stakeholder Groups}
    \label{fig:icorp}
    \Description{This figure gives a summary of our various stakeholder groups and various examples under each group. We describe the groups and examples here: “Part-time & Full-time Wheelchair Users and Advocates” (Including amputees, veterans, athletes, and other manual/power / both wheelchair users); “Primary & Secondary Care Partners” (Including Family Members, Program Directors, and Trip Coordinators); “Park & Recreation Managers”; “Disability Organizations” (Including County level Board of Developmental Disabilities, Residential Services, Student Disability Services, Senior Centers, Retirement Homes, Independent Living, Adaptive Sports Organizations, Veterans Administrations….  ); “Wheelchair Distributors”; “City & Campus Planners” (Including City Councils, Services & Maintenance Directors, Disability Advocates & Experts, Campus Planners & Maintainers…); “Local Businesses” (Including Local Chamber of Commerce, Visitor’s Bureau, Bars, Restaurants, Hotels, Water Parks, Movie Theatres, Concert Halls...)}
\end{figure}

The user interviews were conducted as a part of the National Science Foundation’s Innovative Corps (I-Corps) program \cite{LuberMyPathAnonym} that focuses on interviewing possible users/consumers of a novel technology to validate the idea of the researchers and to evaluate the market demands and requirements. We interviewed 51 individuals from a variety of stakeholder groups which are summarized in Fig.~\ref{fig:icorp}. We have discussed the details of the hypotheses, method, participants, and results of our interviews in Section 4.

Apart from the interviews, we also collected data through a set of Participatory Action Research (PAR) groups. We have formed PAR groups using various experts who are wheelchair users or work with accessibility and community participation aspects. We have a total of 12 PAR members. We discussed the PAR members, the PAR method, and the detailed outcomes in Section 5.

\subsection{Hypotheses for Testing}
\label{sec:hypotheses}
To empirically evaluate our proposed accessibility model we have identified the following hypotheses. The hypotheses were developed by our extensive literature survey and through our experience gained while working with wheelchair users. Each of the hypotheses shows the living experience of daily interactions of users, their wheelchairs, and the environment.
\begin{enumerate}

\item Wheelchair users face greater challenges in finding and navigating routes in unfamiliar places than familiar ones.

\item Wheelchair users often spend a considerable amount of time planning a trip in the community

\item Planning a trip to unknown places incurs more time than to the known places

\item Wheelchair users feel reluctant in community participation unless the location is known to be accessible

\item The difficulty of finding if a destination and the route are accessible emotionally puts off wheelchair users

\item Natural environments like parks and other outdoor public places are usually more difficult to access than the built areas and indoor environment

\item Wheelchair users often find that available community accessibility information for places are negligible

\item Accessibility for wheelchair users is affected by adverse weather conditions like rain and snow.

\item Given all the above challenges wheelchair users can be benefited with the help of a route planning tool  

\item Such a path planning tool/app can be attractive to various other stakeholders apart from solely the wheelchair users

\item Care partners of wheelchair users can use such a tool to plan ahead for an outing with family or community members


\item Community partners like city councils, park \& recreation managers, and visitor's bureau, would be interested in such a tool to mark accessible features of public areas and maintain those features in a timely manner

 
\end{enumerate}

\section{Data Collection through Interviews}

In this section, we have discussed our I-CORP-style interview details. We first describe our participants, followed by the interview method, and then discuss the outcomes of the interviews in three separate subsections. 

\subsection{Interview Participants}

To test all the hypotheses listed in Section~\ref{sec:hypotheses} we have considered a wide variety of participants for conducting our interviews. While Fig.~\ref{fig:icorp} depicts the summary of the stakeholder groups we have considered for our interviews, Table~\ref{tab:part1} shows our 51 participants in more detail. Abbreviations used in Table~\ref{tab:part1} are explained in Table~\ref{tab:part2}. We have a total of 15 wheelchair users which includes manual, power, and power-assist wheelchairs as well as one mobility scooter user. Many users use more than one mobility aid device. Among the interviewees, we have 10 young adults or college students, 11 older adults or retirees, and the rest were middle-aged men or women. Apart from the wheelchair users, we have considered a plethora of other roles (See Fig.~\ref{fig:icorp}) including - Campus Accessibility coordinators (4), Disability Organization employees (7), Senior Living Center Managers (6), City Planners or City Council employees (4), Park and Recreation Managers including city parks and national parks (5), Adaptive Sports Associates (6), Veteran Administration employees (3), Disability Advocates (6), Faculty Members of Disability Studies (6), and some others. The number inside parenthesis stands for the number of persons in that role we interviewed. Many of the interviewees serve multiple roles, e.g., Interviewee P12 (in Table~\ref{tab:part1}) is a Disability Studies faculty (DST) who also serves in the roles of a Caregiver of a wheelchair user (COA), Friend or Advocate of PwD (AD), and a Recreational Therapist (RT).

The major outcome of this research is the identification of various challenges that impact wheelchair accessibility in the community. Building a successful and sustainable solution addressing these challenges would require first-hand experience as well as knowledge from stakeholders who provide these accessible services. Therefore, we have interviewed a diverse field of individuals who connect the greater community for a more holistic approach to modeling accessibility for wheelchair users. The involvement of independent wheelchair users, caretakers, and other stakeholders allows us to test our hypotheses from different perspectives and bring forth novel ideas and aspects regarding them.

\subsection{Interview Method}

We interviewed all 51 participants within a period of 5 weeks. Our primary objective was to test all the hypotheses listed in Section~\ref{sec:hypotheses}. However, we also asked questions to study the demographics of the participants, their level of community engagement, their experience while outing, their comfort level, and interest to use smartphones and various assistive technology solutions, and their view about the need for a path planning tool. Depending on how much the participant wanted to talk, the interviews went on for 15-60 minutes. Due to the risk of the COVID-19 pandemic, the interviews were conducted mostly via Zoom-based video conferencing. However, there were also 5 in-person interviews and 2 telephonic interviews. We have provided a comprehensive list of the interview questions in the associated supplementary file. Since we plan to collect accessible navigational experiences of people spread across various weather and city conditions, we have chosen interviewees from various parts of the USA. Although Table~\ref{tab:part1} shows the primary locations of the interviewees, they have shared experiences of traveling through or living in most of the states across the USA. Apart from that, we have one interviewee from outside the USA who is a caregiver of two wheelchair users in India.


Five of our researchers acted as interviewers. Since our ultimate objective was to develop a route planning tool for wheelchair users, and we wanted to gauge the need for such a tool, to avoid any bias, our interviewers refrained from mentioning our objectives and/or design considerations to the interviewees. Our interviewers focused on the following strategies while interviewing:

\begin{itemize}
    \item Asking the best questions to elicit quick and useful answers related to the hypotheses
    \item Locating and recruiting more interviewees through personal references
    \item Teaching possible ways to conduct the interviews in the most efficient manner for optimal research outcome 
\end{itemize}

Across the three distinct participant groups, it is noteworthy that the only consistently asked questions included those related to demographics and requests for additional contacts or queries. The interviews did not follow a fixed script, and interviewers were encouraged to ask additional questions based on the interviewees' responses. While there was no standardized script, similar questions were posed within each group, with variations across the different groups. In light of this dynamic and adaptive approach to inquiry, a comprehensive analysis of the questions employed across the entirety of the 51 interviews was executed, allowing for the discernment of discernible patterns and variations in the resultant dataset.

\begin{figure}
\begin{center}
\fboxsep=1mm
\fboxrule=4pt
\fbox{\includegraphics[width=0.65\textwidth]{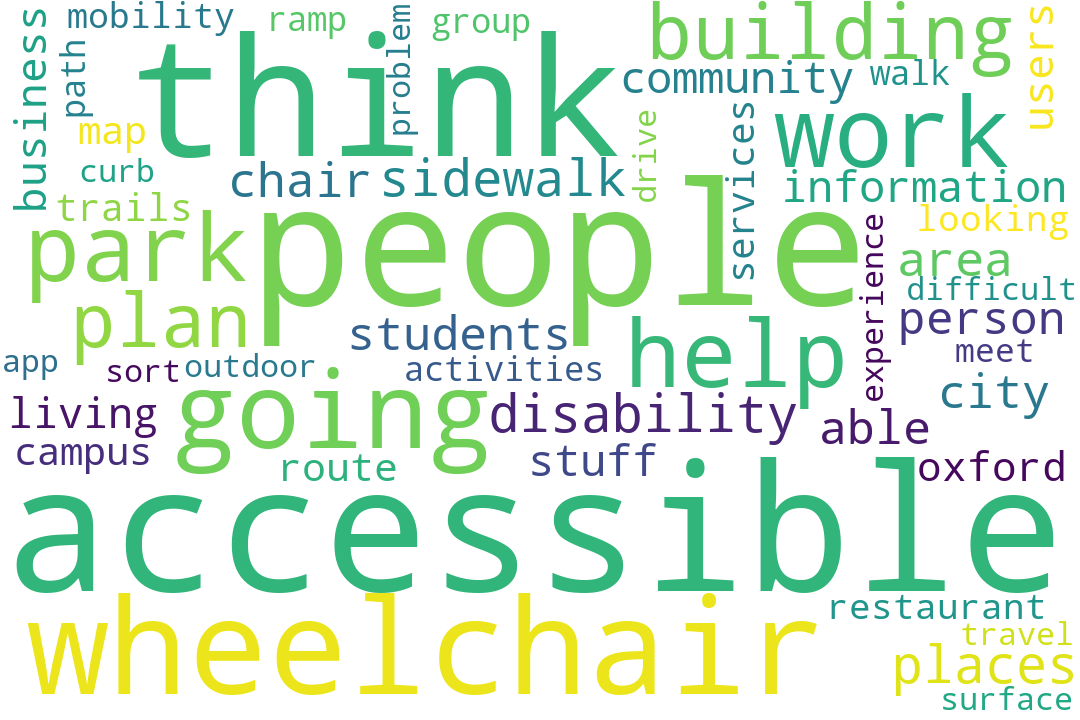}}
    \end{center}
    \caption{Word Cloud Generated from the Transcripts of the Interviews}
    \label{fig:cloud}
    \Description{This figure is a word cloud generated from the complete list of transcripts of the interviews. In a black boundary, we have a figure consisting of words of different sizes and written in different directions. The biggest words represent those that were featured the most in the transcripts. The smaller words are used less but are still greatly used among the transcripts. These words are shown in various colors, many being different shades of green. Some of the highest used words are: ‘Accessible’, ‘people’, ‘wheelchair’, ‘plan’, ‘disability’, etc.}
\end{figure}

Our total interview time was 1383 minutes for 51 participants and it generated transcripts worth 384 pages consisting of 216952 words. Fig.~\ref{fig:cloud} shows a word cloud depicting the important words and themes that came up repeatedly during our interviews. 


We cleaned the transcript and conducted thematic analysis \cite{braun2006using}. Words pertaining to our hypotheses were separately identified and later translated into codes. Participant responses matching or contradicting our hypotheses were thoroughly analyzed before finalizing the conclusive themes. 

\begin{table}[]
\caption{Demographics of Interviewees: This table contains a full list of demographic information for the interviewees involved in the analysis. The headers at the top of the table consist of “Interviewee”, “Age”, “Location”, “Role”, and “Type of Wheelchair”. There are 51 rows listed below the headers providing the results for all 51 interviewees.
}
\label{tab:part1}
\centering
\scalebox{0.9}{
\begin{tabular}{ccccc}
\rowcolor[HTML]{D9D9D9} 
\textbf{Interviewee} & \textbf{Age}             & \textbf{Location} & \textbf{Role}      & \textbf{Type   of Wheelchair} \\
\rowcolor[HTML]{FFFFFF} 
P1                   & 18-25            & Ohio              & CA, FT             & PC                            \\
\rowcolor[HTML]{FFFFFF} 
P2                   & 18-25            & Ohio, Texas              & S, FT              & MC, PC                        \\
\rowcolor[HTML]{FFFFFF} 
P3                   & 18-25            & Ohio              & S, PT              & MC,SC                         \\
\rowcolor[HTML]{FFFFFF} 
P4                   & 26-35             & Ohio              & PR                 & -                             \\
\rowcolor[HTML]{FFFFFF} 
P5                   & 18-25            & Wisconsin         & S, FT              & PC, PA                        \\
\rowcolor[HTML]{FFFFFF} 
P6                   & 26-35             & Tennessee         & WD                 & -                             \\
\rowcolor[HTML]{FFFFFF} 
P7                   & 18-25            & Tennessee, Ohio              & CA, PT             & MC, PC                        \\
\rowcolor[HTML]{FFFFFF} 
P8                   & 26-35             & Ohio              & CPR                 & -                             \\
\rowcolor[HTML]{FFFFFF} 
P9                   & 26-35             & Ohio              & CPR                 & -                             \\
\rowcolor[HTML]{FFFFFF} 
P10                  & 26-35             & Ohio              & CPR                 & -                             \\
\rowcolor[HTML]{FFFFFF} 
P11                  & 26-35             & Ohio              & CPR                 & -                             \\
\rowcolor[HTML]{FFFFFF} 
P12                  & 26-35             & Wisconsin         & COA,   AD, DST, RT & -                             \\
\rowcolor[HTML]{FFFFFF} 
P13                  & 26-35             & Ohio              & SL                 & -                             \\
\rowcolor[HTML]{FFFFFF} 
P14                  & 26-35             & Ohio              & SL                 & -                             \\
\rowcolor[HTML]{FFFFFF} 
P15                  & 36+    & Ohio              & SL                 & -                             \\
\rowcolor[HTML]{FFFFFF} 
P16                  & 26-35             & Wisconsin         & CA                 & -                             \\
\rowcolor[HTML]{FFFFFF} 
P17                  & 36+ & Ohio              & AD,   DST          & -                             \\
\rowcolor[HTML]{FFFFFF} 
P18                  & 36+ & Ohio              & CPR                 & -                             \\
\rowcolor[HTML]{FFFFFF} 
P19                  & 26-35             & Ohio              & PR                 & -                             \\
\rowcolor[HTML]{FFFFFF} 
P20                  & 26-35             & Ohio              & SL                 & -                             \\
\rowcolor[HTML]{FFFFFF} 
P21                  & 36+ & Ohio              & SL,   COA          & -                             \\
\rowcolor[HTML]{FFFFFF} 
P22                  & 18-25            & Ohio              & S, COA             & -                             \\
\rowcolor[HTML]{FFFFFF} 
P23                  & 18-25            & Ohio              & S, COA             & -                             \\
\rowcolor[HTML]{FFFFFF} 
P24                  & 36+ & Ohio              & VA,   AS, FT       & MC                            \\
\rowcolor[HTML]{FFFFFF} 
P25                  & 26-35             & Wisconsin         & VA, AS             & -                             \\
\rowcolor[HTML]{FFFFFF} 
P26                  & 26-35             & Wisconsin         & AS                 & -                             \\
\rowcolor[HTML]{FFFFFF} 
P27                  & 26-35             & Wisconsin         & CPL                & -                             \\
\rowcolor[HTML]{FFFFFF} 
P28                  & 26-35             & Wisconsin         & PR, FT             & PC                            \\
\rowcolor[HTML]{FFFFFF} 
P29                  & 18-25            & Washington        & S, FT              & MC                            \\
\rowcolor[HTML]{FFFFFF} 
P30                  & 26-35             & Wisconsin         & DO                 & -                             \\
\rowcolor[HTML]{FFFFFF} 
P31                  & 26-35             & Wisconsin         & VA,   AS, RT       & -                             \\
\rowcolor[HTML]{FFFFFF} 
P32                  & 26-35             & Wisconsin         & CPL                & -                             \\
\rowcolor[HTML]{FFFFFF} 
P33                  & 26-35             & Ohio              & AD,   DST          & -                             \\
\rowcolor[HTML]{FFFFFF} 
P34                  & 26-35             & Ohio              & DO                 & -                             \\
\rowcolor[HTML]{FFFFFF} 
P35                  & 26-35             & Ohio              & DO                 & -                             \\
\rowcolor[HTML]{FFFFFF} 
P36                  & 26-35             & Ohio              & DST                & -                             \\
\rowcolor[HTML]{FFFFFF} 
P37                  & 36+ & Oregon            & AD                 & -                             \\
\rowcolor[HTML]{FFFFFF} 
P38                  & 18-25            & Wisconsin         & SL, AD             & -                             \\
\rowcolor[HTML]{FFFFFF} 
P39                  & 18-25            & Pennsylvania      & AS, PT             & MC                            \\
\rowcolor[HTML]{FFFFFF} 
P40                  & 36+ & Wisconsin         & DST                & -                             \\
\rowcolor[HTML]{FFFFFF} 
P41                  & 26-35             & Indiana           & FT                 & MC                            \\
\rowcolor[HTML]{FFFFFF} 
P42                  & 26-35             & Maryland          & CA                 & -                             \\
\rowcolor[HTML]{FFFFFF} 
P43                  & 26-35             & India             & AD                 & -                             \\
\rowcolor[HTML]{FFFFFF} 
P44                  & 36+ & California        & AD                 & -                             \\
\rowcolor[HTML]{FFFFFF} 
P45                  & 18-25            & Tennessee         & FT                 & PC                            \\
\rowcolor[HTML]{FFFFFF} 
P46                  & 36+ & Washington        & PR                 & -                             \\
\rowcolor[HTML]{FFFFFF} 
P47                  & 26-35             & Ohio              & DO, FT             & PA                            \\
\rowcolor[HTML]{FFFFFF} 
P48                  & 36+ & Ohio              & DO, PR             & -                             \\
\rowcolor[HTML]{FFFFFF} 
P49                  & 26-35             & Wisconsin         & DO, FT             & PC                            \\
\rowcolor[HTML]{FFFFFF} 
P50                  & 36+ & Wisconsin         & FT                 & PC                            \\
\rowcolor[HTML]{FFFFFF} 
P51                  & 26-35             & Wisconsin         & AS, FT             & MC, PC                       
\end{tabular}
}
\end{table}

\begin{table}[]
\caption{This table is used to explain the different Terms and Abbreviations Used in Table~\ref{tab:part1} and Table~\ref{tab:PAR}. The first group of abbreviations is for the different roles the interviewees hold and have a second column showing the corresponding abbreviation. The second grouping of abbreviations is for the different types of wheelchairs and a second column with the corresponding abbreviations.}
\label{tab:part2}
\centering
\begin{tabular}{cc}
\rowcolor[HTML]{D9D9D9} 
\textbf{Roles}                                      & \textbf{Abbreviations}                        \\
Part-time wheelchair User                            & PT                                           \\
Full-Time wheelchair User                            & FT                                           \\
Caregiver of older adult or relative in a wheelchair & COA                                          \\
Campus Accessibility                                 & CA                                           \\
Disability Organization                              & DO                                           \\
Senior Living                                        & SL                                           \\
City Professional                                    & CPR                                          \\
Parks and Rec                                        & PR                                           \\
Veteran's Admin                                      & VA                                           \\
Blind or Low Vision                                  & BLV                                           \\
Adaptive Sports                                      & AS                                           \\
Campus Planning                                      & CPL                                          \\
Friend or Advocate of PWD                            & AD                                           \\
Disability Studies Faculty                           & DST                                          \\
Student                                              & S                                            \\
Recreational Therapist                               & RT                                           \\
\rowcolor[HTML]{FFFFFF} 
Wheelchair Distributor                               & WD                                           \\
\rowcolor[HTML]{FFFFFF} 
\multicolumn{1}{l}{\cellcolor[HTML]{FFFFFF}}         & \multicolumn{1}{l}{\cellcolor[HTML]{FFFFFF}} \\
\rowcolor[HTML]{D9D9D9} 
\textbf{Wheelchair Type}                             & \textbf{Abbreviation}                        \\
Manual Chair                                         & MC                                           \\
Power Chair                                          & PC                                           \\
Power Assist                                         & PA                                           \\
Scooter                                              & SC                                          
\end{tabular}
\end{table}

\subsection{Interview Outcomes}

In this section, we summarize our findings through the interviews. The first subsection discusses the major themes common to various stakeholder groups while the second subsection verifies if our prior hypotheses hold against the interview responses.

\subsubsection{Major Themes}

We identified a number of underlying themes using the standard methodologies \cite{braun2006using}.

\begin{enumerate}
    \item \textbf{Accessibility Needs Vary from Person-to-Person: } There is no uniform accessibility solution for each person with a disability (PwD) who are wheelchair user for the current research. Some places which may be accessible for one wheelchair user may be completely inaccessible to others. One wheelchair user expresses frustration about planning an outing to a restaurant: ``\textit{I have lost track of how many times I've called to ask if a restaurant is accessible." They respond ``..yes, we are accessible," and then I’ll get there and there's a step, or maybe the front area is accessible, but then you have to go through a narrow hallway to get to the seating, and I can’t do that.}” To able-bodied individuals accessibility is just considered as having an accessible parking spot or a street-level entrance with no stairs. However, the narrow width of an entrance and a hallway, the presence of sharp turns, and the absence of properly accessible restrooms are hardly ever considered. Accessing hilly terrains is also not properly discussed for wheelchair users. Especially for manual wheelchair users, sharp incline itself is fairly difficult let alone steep hills. Businesses may not be aware that not all people with disabilities have the same ability level, so just because one individual can navigate a space or slope, does not mean everyone can, and they can even be putting patron's safety at risk: ``\textit{They said ‘yeah, we have a ramp to get on there.’ Okay, the ramp was literally a 90-degree angle and it took three of us to get the wheelchair down, and then to push it back up, [It was] totally unsafe.}”  Interviewees from Seattle (Washington) and Knoxville (Tennessee) have mentioned hills as a major obstacle to community participation.
    
    \item \textbf{Accessibility of Older Public Spaces: } Many of the interviewees noted that the age of a building or other public areas plays a role in the absence or degradation of accessible infrastructure. Moreover, buildings constructed before 1990 do not conform to the ADA regulations. Some of the older infrastructures are dotted with broken sidewalks and other uneven surfaces; narrow entrances and hallways; absence of curb ramps and accessible restrooms; multiple steps, etc. Participants have witnessed these types of features in Europe and North-eastern cities in the United States. Older cities such as New York, Nashville, and Seattle have been specifically cited by interviewees as inaccessible.\textit{``[New York] was a madhouse … there were places that you'd have to go across another way to avoid hitting a big old pothole or you get run over by people … it's not designed for people in wheelchairs.”} Cobblestones have repeatedly been reported by wheelchair users as the most inconvenient of all surfaces found in older cities:
    However, there are variations in cobblestones ``\textit{...the frequency of vibration or frequency of bumps.. (of).. all cobblestone ..(are not)... equal. So, you might meet cobblestone that's totally fine or you might meet one that feels like you're getting a foot massage because your wheelchairs vibrating so much...}". One interviewee who is the manager of a senior living center clearly mentioned that it is often infeasible to adapt the older buildings to make them accessible: ``\textit{..it's hard because Without building the homes, right from the get-go to be accessible, you are trying to modify older buildings and that's very difficult..}".

    \item \textbf{Pre-outing Accessibility Research by Wheelchair Users: } Interviewees revealed that it is not common to carry out extensive research about the path and destination accessibility before an outing. The destination can be any public space including the built and natural environment, like shopping malls, airports, restaurants, parks, etc. Participants mostly "show up" and decide accordingly if the path and/or destination is not accessible to them. However, for longer trips, they plan better by checking out web-based information, peer user recommendation, mapping tools including street views, etc.

    \item \textbf{Choosing Proper Wheelchair based on Accessibility Information: } In the absence of detailed information about the path and destination accessibility, surface type, slope, etc., wheelchair users cannot decide about the type of wheelchair (manual, power, etc.) they should bring for the trip. One interviewee who otherwise prefers a manual wheelchair because of its lighter frame and easiness of transport finds it unsuitable for some terrains: ``\textit{It’s more getting over the terrain; if I hit a rock in the manual chair, it could flip over because it is so lightweight...}”.

    \item \textbf{Access to Restrooms: } Provision for accessible restrooms is a very important issue among wheelchair users. Older adults are prone to using restrooms more frequently and their care partners find it challenging to locate a usable accessible restroom. Many interviewees reported not being able to access restrooms even if they exist. One individual (a caregiver) mentioned struggling to push older adults in wheelchairs through the stall door. ``\textit{you need to go to the bathroom and it says there's an accessible stall, but in fact it really isn't because they can't wheel their wheelchair in, pivot around get off the wheelchair onto the toilet and get back out … it meets minimum Americans with disability act standards, but it really isn't accessible.}" 

    \item \textbf{Accessibility as Perceived by the City Management: } Individuals that work in City Management positions are aware of the accessibility needs of wheelchair users (PwDs, in general) but are facing some roadblocks while trying to make the community accessible. Due to ADA guidelines and the greater community awareness of accessibility needs, city professionals have implemented a push for community accessibility. Most groups have fixed budgets set aside either for general community accessibility, or more specific accessibility projects, such as installing greater amounts of curb cuts. Many of the interviewees addressed that the funds are what make the projects possible, however, insufficient funds can prevent an accessibility project to take place. Another setback to community accessibility is how time-consuming the projects can be. Many reported accessibility issues are large-scale and may take several years to resolve, especially if there are limited funds for the project itself. One interviewee who is a city manager mentioned ``\textit{Every year we've invested over $100,000$ for the past 15 years or so in adding ADA compliant ... ramps to the sidewalk system, and we believe we will have federal funding to finish that project and have an ADA ramp at every sidewalk corner within the next year.}”. 

    \item \textbf{Gathering City Accessibility Information through Crowd-sourcing: } Many employees of the city management do not have accessibility needs, so their perception of what the community needs is often based on what is reported directly to them. Many of our interviewees spoke of seeking out help from student groups, advocacy organizations, and community members with accessibility needs. Since these workers oversee many aspects of the community, they do not always know all of the accessibility barriers that are present if they are not directly reported to their office. Accessibility is also a broad term and is very person-specific. This often overwhelms city management in prioritizing one accessibility need over the others and addressing them. One interviewee even stated ``\textit{it just becomes overwhelming ... I just throw my hands up in the air, because it becomes very daunting of where to start.}”

    \item \textbf{Limited Accessibility of Natural Areas for Wheelchair Users: } We interviewed Parks and Recreation managers of city councils and wheelchair users accessing natural areas and found various useful themes. One common finding is that park accessibility exists but has limits. Due to ADA guidelines and popular interest in park accessibility, there has been an increase in the information available to wheelchair users both online and onsite. Interviewees reported parks having accessibility information posted onsite on trail notices and some parks have online information about accessibility to help in prior planning. Also, it was reported that there is a limit to how far accessible areas reach as people do not want to make many changes to the natural environment. Interviewees stated not being able to go more than a few miles in a park due to the accessible trails and sidewalks ending. A park and recreation manager noted the difficulty of making the outdoors accessible beyond a certain point. They accept that ``\textit{there is no way to control the outdoors, there are too many variables involved.}" 
    
    \item \textbf{Participants Prefer to Choose the Extent of their Outdoor Activity: } We found out that some wheelchair users do not need a lot of accessibility options as they would rather figure out how to adapt on their own terms. Wheelchair users who are passionate about being outdoors like to accommodate nature and find ways to do the things they want to do. One wheelchair user mentions ``\textit{Mother nature doesn’t have to conform; we have to conform.}” Also, they prefer to choose their level of participation: ``\textit{The level of accessibility you need in order to feel comfortable and to be willing to participate... Choose the level of challenge that you want to engage in}”. Interview outcomes show that park and recreation situations for wheelchair users depend greatly on the user's mindset. If an individual has no interest in the outdoors, they might struggle more when attempting to participate. On the other hand, if someone is passionate about the outdoors, they will not let inaccessibility hold them back from doing what they want to do. They will find a way to make their activities work for them. There is a balance that needs to be found between making accommodations and feeling comfortable enough to participate.

    \item \textbf{Views of Disability Organizations on Community Accessibility:  } We interviewed persons associated with various disability organizations and recorded their views. Regarding community accessibility, the quality of the sidewalks, the presence of curb cuts, access to ramps, and physical barriers are all components of which these organizations must be aware. Many of the interviewees mentioned that access to the most up-to-date information about these areas can ensure the individuals they are serving can have the safest, independent, and fulfilling experiences in the community.
    
\end{enumerate}

\subsubsection{Hypotheses Confirmation}

Once we identified the key themes arising from the interviews, we started investigating their matches with the hypotheses listed in Section~\ref{sec:hypotheses}. We can see that of the ten major themes, seven match partially or completely with our original hypotheses. The other three themes emerged as new insights compared to our previous hypotheses.

We did not hypothesize any relations of accessibility with the older city infrastructure as found through theme (2). Through our interviews, we discovered that wheelchair users often face more challenges navigating older cities which have hardly any accessible infrastructure. They often require spending more time and using additional tools in planning outings to those places. Some wheelchair users noted that southern states are more accessible than northern ones. A veteran athlete in a wheelchair (P24) who frequently moves around the USA mentioned: ``\textit{Arizona is very wheelchair friendly. a lot of the cities down South, little rock, the areas down south are more apt for wheelchairs, because a lot of the people from the north, the older ones go down there for the winter}." And at least 2 more wheelchair users informed us that southern states having warm weather want to attract retirees and vacationers who are older adults and have better accessibility features. However, another wheelchair user (P45) does not think accessibility has anything to do with location but how old the city is (as captured above in theme (2)): ``\textit{...accessibility has had less to do with the geographical location. And more to do with how old the city is, and I think the in the north there are a lot more older cities and right along the coast things are very old because it's kind of the first settlements..}."

Finding accessible restrooms in the destination location or nearby came up as a recurring theme across many of our interviews although overlooked during our hypothesizing. It appeared to be very important for a successful outing for wheelchair users and often forces them or their caregivers/tour planners in researching nearby restroom locations.

Another missing hypothesis that appeared in theme (7) is about employing crowd-sourcing to receive knowledge about path barriers from the community. Many interviewees were generally content with their current routing system but expressed interest in innovative ways to provide and receive more feedback regarding accessibility barriers, and various ways to access related information to facilitate community accessibility. 

Our hypothesis (2) regarding thorough route planning by wheelchair users prior to community participation did not fully align with the experience of the interviewees (see Theme (3)). Multiple interviewees clearly mentioned that they refrain from pre-planning every trip they take in familiar or nearby areas. However, for long-distance trips to unfamiliar spaces, they plan ahead and prepare accordingly which confirms the hypothesis (2). Specifically, disability support organizations that plan events and community outings with wheelchair users specified researching, calling ahead, location scouting, and "dry runs" prior to an actual outing. ``\textit{it's all about the scoping you know we definitely don't do an event without sending a committee or at least one or two people to do a dry run}”

In our hypothesis (8) we mentioned the effect of adverse weather on wheelchair accessibility. Although the interviewees agreed with the hypothesis, they brought up a novel perspective that we did not hypothesize. Wheelchair users mentioned that they decide on the type of wheelchair (manual, power, etc.) to use depending on the weather condition. If the weather and terrain are not known prior, users prefer to use the device that can perform most reliably instead of their preferred wheelchair. ``\textit{I use both an electric and a manual depending on the weather, if it’s raining, I tend to use my electric chair.}"

We hypothesized in (5) that an inaccessible path may thwart a wheelchair user. Our interviewees have confirmed this hypothesis in general with some extreme cases. Interviewee P2 (in Table~\ref{tab:part1}) who is a mobility scooter user needed to change out of his/her favorite major.

``\textit{I was forced to switch from my dream major because ...the ...building was ...notorious .... the hallways were extremely narrow with sharp turn so ... you couldn't get around corners in a wheelchair or a scooter... there was an elevator but it was absolutely minuscule, I cannot fit a wheelchair or a scooter. The floors were uneven .. and there were just so many ... steps (in)...the entire building I just simply cannot ...get anywhere within that even if I could get inside.}"

Another similar experience by P45 made him/her stuck in the outdoors for a very long time on a winter evening: ``\textit{I ended up stuck on campus in the cold. ..I think it was the beginning of October .. and I wasn't able to get home until 12:30 AM. ... I was calling everybody I could think of ... saying, is there just an accessible vehicle that can get me home? And they said they could send an ambulance. And .. that's not at all what I need, ... one bus driver overheard and out of kindness ... came and ... brought me home even though it was very much against the rules for him. You know he wasn't supposed to go off the route. .. I ended up really ill after that for a couple of weeks. I had bad ear infections because I was out in the cold the whole evening.}"


\color{black}

Analysis of our interview outcomes bolsters the need for a path planning tool for wheelchair users which will incorporate accessibility information and help in pre-outing planning with relative ease.

\section{Data Collection through Participatory Action Research (PAR)}

Apart from the interviews we also collected data through action research and used those to validate our proposed hypotheses (see Section~\ref{sec:hypotheses}). in the following sections, we discuss details of our PAR members, the PAR method, and the results of the action research.

\subsection{PAR Members}
We planned to recruit a diverse team of PAR members who can provide unique perspectives about accessible wheelchair routing. We specifically focused on recruiting individuals with different types of disabilities such as developmental, visual, and ambulatory. Some of our PAR members are individuals who work with wheelchair users on accessibility issues, such as university campus planners, city accessibility coordinators, and various disability organizations. Recruitment was done by contacting via email and phone. Table~\ref{tab:PAR} shows the list of our PAR members and their roles (See Table~\ref{tab:part2} for roles and their abbreviations).

\begin{table}[H]
\caption{Demographics of PAR Members: This table shows demographic information for the 12 PAR members involved in the Participatory Action Research. The headers of the table consist of “PAR Member”, “Age”, “Location”, “Role”, and “Type of Wheelchair”. The table has 12 following rows providing the details for the PAR members.}
\label{tab:PAR}
\centering
\begin{tabular}{ccccc}
\rowcolor[HTML]{D0CECE} 
PAR Member & Age         & Location  & Role                            & Type of Wheelchair \\
PAR1         & 26-35  & Wisconsin & \cellcolor[HTML]{FFFFFF}CPR, DO & -                  \\
PAR2         & 26-35  & Ohio      & \cellcolor[HTML]{FFFFFF}CPL     & -                  \\
PAR3         & 18-25 & Wisconsin & FT, S                           & PC                 \\
PAR4         & 26-35  & Tennessee & \cellcolor[HTML]{FFFFFF}FT, S   & PA                 \\
PAR5         & 18-25 & Ohio      & PT, DO                          & PA, MC             \\
PAR6         & 26-35  & Ohio      & SL, DST                         & -                  \\
PAR7         & 36-50 & Ohio      & BLV, DO, AD                      & -                  \\
PAR8         & 36-50 & Wisconsin & BLV, AD                          & -                  \\
PAR9         & 26-35  & Wisconsin & FT, AD                          & MC, PA               \\
PAR10        & 26-35  & Maryland & CA, AD                          & -                   \\
PAR11        & 36-50  & Wisconsin & DO                           & - 
              \\
PAR12        & 26-35  & Tennessee & WD                          & -    
\end{tabular}
\end{table}

\subsection{PAR Methods}

Participatory Action Research (PAR) is a collaborative approach to collecting information about an issue involving a group of people who are somehow involved in that issue and/or are concerned about the same \cite{kindon2007participatory}. PAR members 'participate' in a discussion, identify the challenges they face, and provide solutions through a series of repetitive steps involving `Action' and `Reflection'. Once our PAR members were identified we presented in front of them our initial set of hypotheses and involved them in an in-depth discussion process. We obtained their feedback in terms of their personal experiences in using a wheelchair or assisting other wheelchair users.


\subsection{PAR Results}

PAR members discussed our proposed set of hypotheses introduced in Section~\ref{sec:hypotheses} and 
identified various `\textit{Issues}' and provided thoughts (`\textit{Reflections}') to tackle those issues. We summarize them below.
 
\begin{itemize}
    \item \textbf{Issue}: Non-uniform access to technology. \textbf{Reflection}: For older adults, technology access can be challenging. Also, technology access depends on racial components and socioeconomic status. It was recommended to include people from every racial group and socioeconomic status to avoid implicit bias in obtainable data.

     \item \textbf{Issue}: Capturing weather and environmental changes are non-trivial. \textbf{Reflection}: Rapid changes in weather and environment are common. They may lead to worsening surface conditions. Capturing those sudden changes is necessary to build a path planning tool. A crowd-sourcing-based method can be useful.

    \item \textbf{Issue}: Difficulty of using multiple devices. \textbf{Reflection}: Since the path planning tool will be required to provide turn-by-turn navigation and wheelchairs often do not have smartphone holders, using a phone and a wheelchair at the same time can be very difficult for wheelchair users. This issue can be addressed by including voice commands and turn-by-turn voice directions.  

    \item \textbf{Issue}: Different wheelchairs differ in characteristics. \textbf{Reflection}: It is important to discern between different types of wheelchairs - manual, power, power-assist and their variations. Different types of wheelchairs produce different vibration patterns given their characteristic differences. A path planning tool must take into account the nature of the wheelchair and its various features to find an accessible route.

    \item \textbf{Issue}: Populating an accessibility map with a critical amount of data. \textbf{Reflection}: Crowd-sourcing-based systems often fall prey to the data gap where not enough data is there to attract the interest of users. It is imperative to populate information about temporary and permanent barriers in an accessibility map leveraging community knowledge. Also, users bring more users. So, such a path-planning tool should be advertised through public forums, websites, wheelchair builders \& distributors. Furthermore, it is important to locate accessibility “deserts” where people are not accessing areas a lot. This is an opportunity for people to engage in community areas they may not be as familiar with. Also, this urges people to think about how accessible a particular surface might be for others, e.g., the elderly community. And being sensible is the first step to solving a problem.

    \item \textbf{Issue}: Competing with the existing mapping tools. \textbf{Reflection}: While existing mapping tools like Google maps, Apple maps, and Waze are widely used, they lack accessibility information. A novel path planning tool should include a sustainable economic model and consider complementary features to the current mapping tools. This may lead to eventual integration with Google maps, Apple maps, and Waze-like tools. 


    \item \textbf{Issue}: Presence of permanent and temporary barriers. \textbf{Reflection}: A reoccurring issue with mapping tools is the inability to report barriers strewn across paths that render them inaccessible. Temporary barriers such as broken sidewalks, piled-up snow, discarded furniture, etc. do not allow easy transit for wheelchair users. Sometimes, navigation tools may be unable to recognize permanent barriers like stairs and steep slopes which make the path completely inaccessible and the users' journey futile. 
    
    \item \textbf{Issue}: Multi-accessibility of existing routing systems. \textbf{Reflection}: Accessible mapping and navigation systems need to be accessible not only to wheelchair users but also to blind and low-vision, users who are hard of hearing and others. It is crucial for accessible systems to have visual and audio features to assist all users in finding accessible paths.

\end{itemize}


\section{3D U-W-E Accessibility Model}

\begin{figure}
\begin{center}
  \includegraphics[width=0.6\textwidth]{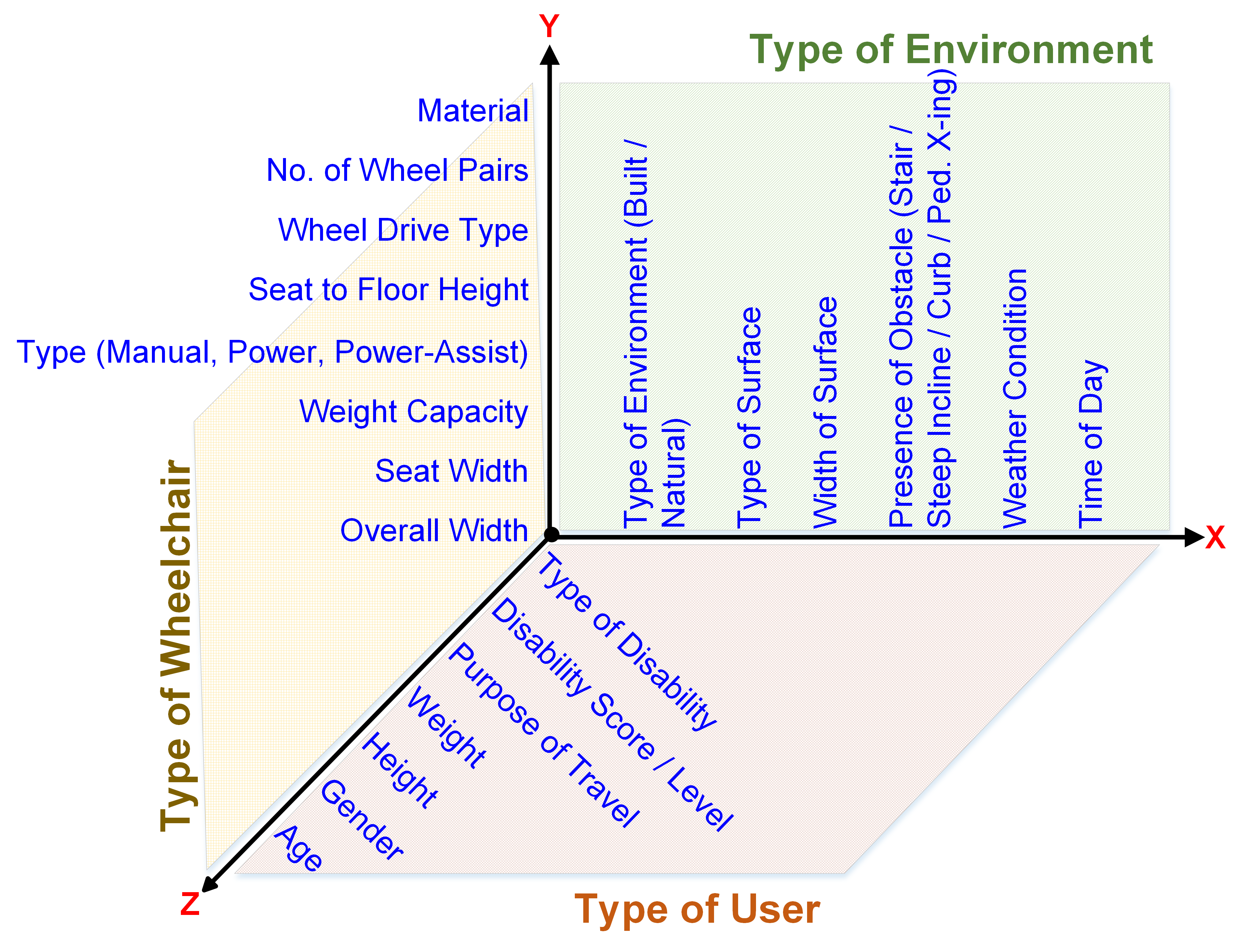}
    \end{center}
    \caption{3D-UWE Accessibility Model}
    \Description{This figure has three different orthogonal dimensions that show “Type of User”, “Type of Wheelchair” and “Type of Environment” as Z-, Y-, and X- axes of a three-dimensional system. Along the dimensions, we list multiple parameters related to User, Wheelchair, and Environment. The values assumed by those parameters can help determine the accessibility of a path. The three orthogonal axes of the figure suggest that different values of the parameters along an axis will not affect the choice of values of parameters along the other two axes. We have identified multiple parameters that can identify the type of user including – age, weight, height, gender, type of disabilities, disability level, and purpose of travel (urgent/ non-urgent). The type of wheelchair can be dependent upon the Manual/Power/Power-assist wheelchairs, the material used, number of wheel pairs, wheel drive type, seat to floor height, seat width, weight capacity, and overall width of the wheelchair. The type of environment can be built/natural, surface type, surface width, presence of obstacles (like a curb, stairs, steep incline, pedestrian crossing), weather conditions, like sunny, cloudy, rainy, snowy; time of the day, and purpose of travel.}
    \label{fig:3DUWE}
\end{figure}

Analysis of the outcomes of 51 interviews and the action research discussions helped us to properly understand the interaction of various factors of the accessibility model identified in Section 3. We have noticed that the User (U), wheelchair (W), and the environment (E) - the three different characteristic factors are orthogonal to each other, i.e., choosing a particular characteristic factor does not influence the choice of another characteristic factor. So, we combine the factors listed in Table~\ref{tab:Characteristics} and summarize them using a three-dimensional model of accessibility (3D-UWM) shown in Fig.~\ref{fig:3DUWE}.

Different axes show different dimensions - U, W, and E, and the parameters associated with them. Parameters have no order of appearance within a particular dimension. Although the results were published in other venues \cite{gani-accepted-a} and we have not mentioned that in this paper, we also carried out a quantitative evaluation to verify the impact of various parameters on accessibility. We measured accessibility as a function of the vibration produced by wheelchair movement through a particular surface. Vibration is captured through inertial sensors like accelerometers and gyroscopes. We observed that the vibration changes mostly take place with the change of wheelchairs - e.g., from manual to power wheelchairs.  Overall, our 3D-UWM model is the first ever accessibility model verified qualitatively and quantitatively and will be used in the future for further experimentation regarding wheelchair accessibility.

\section{Discussion and Conclusion}

The accessibility of wheelchair users is often limited by various permanent and temporary barriers in the built and natural environment. This is a well-reckoned problem and researchers have investigated various aspects of the challenges faced by wheelchair users. However, identifying the various dimensions of wheelchair accessibility with respect to users, wheelchairs, and the environment has not been thoroughly investigated. 

In this research, we have involved wheelchair users and various other individuals working to facilitate wheelchair accessibility to identify various issues not considered before. We have conducted 51 interviews and 2 PAR meetings with various stakeholders and verified a set of hypotheses proposed by us as a result of an in-depth literature study. Our thematic analysis and action research helped to both confirm and contradict hypotheses and directed us towards further growth. While we could successfully verify most of our hypotheses, a few different views arose from the interviews and from action research outcomes. 

We notice that there is a great amount of variability in community accessibility needs even among wheelchair users. Even though some wheelchair users do not plan for an outing and carry on with the "show up and deal" attitude, it is not common for all and definitely not for longer trips. Caregivers and trip planners in disability organizations reported conducting extensive planning, and sometimes sending out scouts, prior to every outing. This includes planning every possible scenario including restroom stops. Sometimes wheelchair users refrain from community outings as they do not find anyone to assist them and they are not confident to navigate the path barriers by themselves. Even finding in-home care is difficult for older people.


The outcomes of our study clearly suggest that lack of accessibility discourages people from community participation and a path planning tool may boost the self-reliance of wheelchair users and encourage them to participate more in community activities. Our research team is critically analyzing the findings and trying to incorporate them into the design process of our path-planning tool.

\begin{acks}
This work is partially funded by the National Institute on Disability, Independent Living, and Rehabilitation Research (NIDILRR grant number 90IFDV0024), the Administration for Community Living (ACL), Department of Health and Human Services (HHS).  The content of this work does not necessarily represent the views or policy of NIDILRR, ACL, HHS. For an overview of prior project work and the technical aspects of the project, please visit: http://www.routemypath.com/. The research is covered by Miami University IRB Protocol 0738r and University of Wisconsin Milwaukee IRB 22.264-UWM.

\end{acks}

\bibliographystyle{ACM-Reference-Format}
\bibliography{sample-base}

\appendix

\end{document}